# Context-aware adaptation for group communication support applications with dynamic architecture

Ismael Bouassida Rodriguez[1,2,3], Khalil Drira[1], Christophe Chassot[1,2] and Mohamed Jmaiel[3]

[1] LAAS-CNRS, University of Toulouse, [2] INSA 7 avenue du Colonel Roche, Toulouse, France
[3] Redcad, Enis, Route de la Soukra Sfax, Tunisia
{bouassida, khalil, chassot}@laas.fr, mohamed.jmaiel@enis.rnu.tn

**Abstract**: In this paper, we propose a refinement-based adaptation approach for the architecture of distributed group communication support applications. Unlike most of previous works, our approach reaches implementable, context-aware and dynamically adaptable architectures. To model the context, we manage simultaneously four parameters that influence Qos provided by the application. These parameters are: the available bandwidth, the exchanged data communication priority, the energy level and the available memory for processing. These parameters make it possible to refine the choice between the various architectural configurations when passing from a given abstraction level to the lower level which implements it. Our approach allows the importance degree associated with each parameter to be adapted dynamically. To implement adaptation, we switch between the various configurations of the same level, and we modify the state of the entities of a given configuration when necessary. We adopt the direct and mediated Producer-Consumer architectural styles and graphs for architecture modelling. In order to validate our approach we elaborate a simulation model.

## 1. Introduction

Our work aims to makeing context-aware applications self-adaptable. For this kind of applications, it is necessary to be able to dynamically adapt the architecture during execution. Designing and implementing self-adaptive communicating systems is a complex task, which may be addressed via model-based design approaches associated with automated management techniques for dynamic architectural adaptability. In self-adaptable applications components are created, and connected, or removed and disconnected during the execution. The architectural changes respond to constraints in the communication and resources and to energy variation. They may also respond to evolution in the supported activities. The changes can also result from the user mobility. Static architectures are described by instances of components and interconnection links. This approach appears inappropriate when the architecture structure changes need to be described. The dynamic architectures character requires additional description rules. Several works have addressed the dynamic architecture description, using different approaches [3]. In order to guarantee the architecture updates correctness we use formal techniques. In particular, graphs represent a powerful expressive mean to specify respectively static and dynamic architectures aspects. For architecture description we use ACG (Abstract Component Graph) [10] approach. The graph nodes represent the software components, and the edges represent the links between these components. The evolution is implemented by an architecture transformation specified by a rewriting rule. In this context we are interested in group communication support applications. These applications require adaptation ability to their various contexts. We use direct and mediated Producer-Consumer architectural styles [7]. To illustrate the proposed models and their transformations, we consider a case study of Emergency Management Activities (EMA) involving several cooperating participants or nodes having different roles and functions dealing with several context changes.

This paper, is organized as follows. In section 2, we present the related work. In section 3 we detail our refinement approach for adaptation illustrated on the considered case study. Then, in section 4, we propose a simulation for the refinement approach where we handle the internal structure of nodes and the communication links between the nodes. Finally, in section 5 we conclude.

## 2. Related work

The adaptation solutions suggested in the literature are defined in various ways. Adaptation may be ruled by architecture based transformation laws. In [9] the classification of common adaptation techniques are identified and classified. The adaptation is architectural when the structure of adaptive services can be modified. [9, 8] provides frameworks for designing Transport level protocols whose internal structure can be modified according to the application requirements and network constraints. The replacement of a processing module by another(s) can implement adaptation actions, following a plug and play approach where the new component has the same interfaces as the replaced one. The applications have requirements and constraints provided by entities exchanging high level information. Requirements and constraints may change dynamically depending on the supported cooperative activity and its evolution. Two different adaptability approaches may be distinguished: the design time adaptability and the run time adaptability. For the design time adaptability, we can find commercial tools for application architecture adaptability like AAA of HP [2], a design support tool. The AAA tool handles the application development cycle and optimizes the resource value, by insuring that the infrastructure answering clearly and in a measurable way to activity requirements. For the run time adaptability [9] presents several adaptation techniques among which use proxy services, change model of interaction and reorganize application structure. Adaptability is also implemented by multimedia retrieval systems like the search engine framework CARSA [4]. Behavioral and architectural adaptability are the two approaches addresses by the proposed solutions. [6] describes a distributed image visualization application that consists of components that implement different compression methods. On the other side, at the application level, [13] addresses the need for adaptation in video streaming applications distributed over the Best-Effort. In general the application adaptability is implemented as a handling of features provided by lower layers. The adaptation can also be managed at the component-to-component communication level, aiming at supporting a given application level architecture and considering resource related constraints. Multiple architectures may be designed by adopting partially or totally centralized and distributed styles [9, 11]. Different criteria may be used for routing and connecting strategies. Similarly to [5], we can consider typed channel managers where producers and consumers are connected to one or another manager depending on the type of information they produce or consume. The Peer to Peer technology [12] has initiated publishing and dynamic discovery of components. Service-oriented technologies as Web Services constitute a continuation towards this direction. Service-Oriented Architectures (SOA) are based on dynamically publishing and discovering services. This kind of architectures provides the possibilities to dynamilically compose services for adapting applications to contexts. Service descriptions are published, via the registry, by service providers and dynamically discovered by service requesters. We distinguish also some projects that focus on the middleware-level adaptation like "The Adaptive Application Project" [1].The goal of "The Adaptive Application Project" is to provide a programming framework (a programming model, language, compiler, and runtime environment) that enable programmers to design, develop, and optimize the performance of adaptive distributed applications. Here we can say that the adaptation question is widely addressed in the literature, we distinguish two different approaches: design time adaptability and run time adaptability where we find several adaptation techniques. This diversity is extended here by answering to questions dealing with energy management and computing load balancing.

# 3 Adaptation approach

Managing architectural adaptation requires considering abstraction levels dedicated to specific parts of the whole adaptation, and self-adaptation has to be managed in a coordinated manner both within and between these abstraction levels. Distinguishing these abstraction levels allows designers and developers to master specification and implementation of adaptation rules. For a given configuration $A_{n,1}$ at level n, multiple configurations $(A_{n-1,1}, ,A_{n-1,p})$ may be implemented at level n-1. Adapting the architecture to constraint and requirement changes at level n-1 by switching among these configurations allows maintaining unchanged the n-level configuration. Likewise, when adaptation requires changes at level n, this may need no change at level n-1 if initial and new configurations of level n (e.g. changes from $A_{n,1}$ to $A_{n,2}$) have common implementations at level n-1 (e.g. $A_{n-1,p}$). We consider distributed component-based applications deployed on mobile communication nodes. The communication has to be maintained adapted to the context change factors. These factors are given according to the application and the node properties. Being aware of these factors, that we call context, provides a certain form of adaptability. The application and the node properties are: The mobile nodes move in a limited perimeter, each node has limited resources in term of energy and memory, priorities are associated with the communications among nodes. Moreover, on the same links it is possible to have several types of data with different priority degrees. We drive the evolutions between levels by considering the context changes.

## 3.1 Case study

Recent advances in computing and networking technologies enabled the deployment of complex group communication activities, such as Emergency Management Activities (EMA) involving mobile users cooperating within a common mission. This section presents the graph-based models of the three abstraction levels through a case study related to EMA-like activities.

### 3.1.1 General description

We assume an EMA group composed of a fixed controller, say $A_1$, and two investigators, say $A_2$ and $A_3$, moving within the exploration field. For simplifying the model explanation, coordinator and controller are merged into a single role: controller. Functions performed by investigators include Observing the exploration field and Reporting feedbacks to the controller. Two kinds of feedbacks are distinguished: feedbacks D are Descriptive data; they are transmitted by means of audio/video; feedbacks P are Produced data; they express the analysis of the situation by an investigator. They are transmitted by means of audio. Investigators $A_2$ and $A_3$ provide continuous feedbacks D to $A_1$; they also provide periodical feedbacks P; there is no priority difference between communications $A_2$-$A_1$ and $A_3$-$A_1$, but transmission of feedbacks D is more important than those of feedbacks.

### 3.1.2 Architecture Modelling

The architecture is represented by a directed graph, the vertices represent communication devices hosted by mobile nodes. The edges are labelled by the sent data types from a node to another and the priority of each type. The edge direction indicates the data flow direction. The producer and consumer are respectively at the tail and the head of the arrows. We also specify the factors the architecture has to adapt. This consists in defining the context elements. The elements we consider are: the energy level of the nodes, the memory saturation level of the nodes and the bandwidth available on the link. These three factors and the communications priority degree change during time and trigger the architecture transformations to adapt the application to these changes.

### 3.2 Architectural Refinement

We describe a refinement approach which implements the initial architecture on an event-based architecture, and on which we act for the adaptability. We proceed step by step. Initially we handle the internal structure of each node and then we refine the communications links between the nodes.

#### 3.2.1 The nodes internal structure processing

We also proceed here step by step. Step 1 implements Functional entities decomposition. We consider now, the case of a node ($A_1$). We split the entities inside the node according to their functional role. In our example, we consider that each node has a communication entity C and several processing entities T. At the end of this step, it is possible to proceed in two manners: Mediated Producer-Consumer style and Direct Producer-Consumer style. We detail first, the reasoning for the mediated Producer- Consumer style. Step 2 addresses communication message filtering. The communication entity receives a number of messages addressed to the different processing entities. The event server has the role of an event dispatcher. In one node, we distinguish one communication entity (C) and two processing entities ($T_D$ and $T_P$). Each entity manages one flow type (D or P). Step 3 addresses the communications type choice. We consider the communication between the processing entities and the communication entity inside each node. The D data type has a high priority, therefore we choose a push link. A push link makes it possible to transfer the data as soon as it is produced. The P data type has a low priority, therefore we choose a pull link. A pull link makes it possible to transfer the data on demand of the consumers. We consider the internal structure of a refined producer node. The associated processing entity $T_D$ pushes the data when it is produced. On the other hand, the associated processing entity $T_P$ pulls the data when it is requested. We consider the internal structure of a refined consumer node. The communication entity C pushes the D data type because it has a higher priority. We detail now, the logic reasoning for the direct Producer-Consumer style. Like for step 2 when we consider the mediated Producer-Consumer style, we choose a push link which makes it possible to transfer the D data type as soon as it is produced. We choose a pull link which makes it possible to transfer the data on the consumer demand. In direct Producer-Consumer style, we do not use a filtering entity (event dispatcher). We have direct communication links between the communication entity and the processing entities. The functional differentiation makes it possible to deactivate the least important processing entities when the energy level of the nodes goes down.

#### 3.2.2 Communication links refinement

The communication link refinement is related to the current context. To be able to adapt the system, we associate to each link a "rate". The rate represents the data transfer frequency by a node on a link. Through the rate and the nature of the link we will act to adapt the system. We also define thresholds associated with each parameter the context to define the rates witch trigger the adaptability actions. For instance, we define the scale for the energy level. If the node energy level is between 80% and 100%, the required rate is PR1. If the node energy level is between 40% and 80%, the required rate is PR2. If the node energy level is between 0% and 20%, the required rate is PR3. This rule is defined at the system deployment time and can be modified during the execution automatically or through an administrator intervention. Similar rules are associated with the bandwidth available on the link and the memory saturation level on the node. For each link, we associate a priority degree which is initially set to zero. Each factor can require a specific rate and we define a global link rate. We adopt the following policy. For two nodes A and B, each node calculates the rate required according the threshold rules. We have for each node: ER the rate associated with the energy level of the node; MR the rate associated with the memory saturation level of the node; BR the rate associated with the bandwidth available on the link and LR the rate associated with the link priority degree. Each node calculates his global link rate (GR) as follows: $\frac{a\ ER\ +\ b\ MR\ +\ g\ BR\ +\ m\ LR}{a\ +\ b\ +\ g\ +\ m}$ . Where the values α, β, γ and μ are weights that allow an importance degree to be associated with each factor. For instance, if the administrator knows that for a specific node the memory saturation level is the most important factor, he/she sets β to a value higher than α, γ and μ. The threshold scales for the four factors and the weights constitute the node profile. This profile depends on the technical characteristics of the node in the deployed application. Node A and Node B calculate their global link rates ($GR_A$, $GR_B$). Then the link rate LR is calculated. The new LR corresponds to the minimum between the $GR_A$ and the $GR_B$. We obtain newLR= minimum ($GR_A$, $GR_B$). At the end of the operation, we reach a tradeoff which provides the adaptability until the next context change.

## 4 Simulation model and results

To validate our approach, we simulate the behaviour of an architecture composed of four mobile nodes. The D data type has a high priority and the P data type has low priority. We apply the three refinement steps and we adopt the direct Producer-Consumer style. Each link is mapped on a push or a pull interaction link. The pulls implement the transfer of data P because this transfer has a low priority. The pushes implement transfer of the data D because this transfer has a high priority. We consider the internal structure of each node, our approach allows to deactivate the component that the node does not use. We have two types of processing components ($T_D$ and $T_P$). In each node we represent only the active components. For node $A_4$, we deactivate $T_D$ because this node does not process D data type. For the other nodes, we have the communication entity (C) and two processing entities ($T_D$ and $T_P$). The links between the two processing entities and the communication entity are of type push or pull. We associate the pull interaction mode with the transfer of data P. And we associate the push interaction mode with the transfer of the data D. Each node has a profile: a rate rule and four factor importance values α, β, γ and μ.

### 4.1 Simulation parameters

To simplify, we focus here, on the energy evolution for each node. We consider two random variables: X that represents a

Poisson law of parameter λ, which characterizes the node's message production and Y that represents a Gaussian law with parameters m and σ, which characterizes the bandwidth variation on the links. For the memory state, we use a trace of a mobile node previously logged and we take into account the variable X. We calculate the node energy according to the node state (consuming or producing, inactivity and idle).

### 4.2 Results and interpretations

We give here the curves of node energy evolution during time with and without adaptability. We draw the curves that show the energy evolution on each node with and without adaptability.

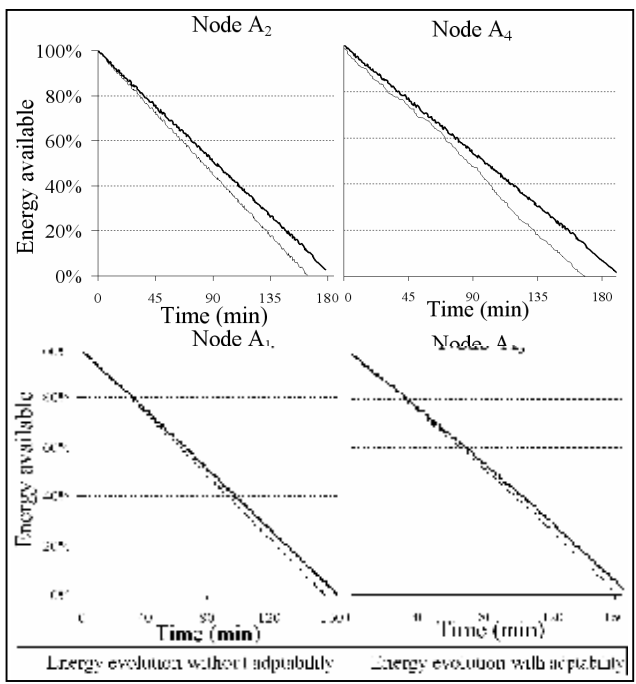

Fig 1. Energy evolution on nodes

We estimate how much time the node stays "alive" or has enough energy to work properly. For node $A_4$ (fig 1), we notice that adaptability provides energy 23 minutes (14%) more. This is due mainly to the deactivation of the processing entity $T_D$ and the fact that node $A_4$ makes only pulls. For node $A_2$ (fig 1), we notice that adaptability provides energy 19 minutes (11%) more. This is due mainly to the adaptation of the push rate. During simulation, the energy decrease made the node decrease the push rate. For nodes $A_1$ and $A_3$ (fig 1), the improvement was only of 8 minutes (5%) and can be improved by a different tuning of the rate rules and the parameters α, β, γ and μ.

## 5 Conclusion and perspective

We presented in this paper an approach of adaptation for group communication support application with context-aware architectures sensitive to the context. We consider different refinement steps that make it possible to decompose the nodes into processing and communication entities. It also allows us to act on the communication links. In particular, we can tune the rate on a link. This adaptation is not static, but it is specific to each node through, the weights α, β, γ and μ and the threshold scales according to the role of the node. To validate our approach, we provided a model to simulate the architectural quantitative attributes. The effectiveness of our approach was shown but we still work on adjusting the importance parameters α, β, γ and μ and the threshold scales. We plan in addition, to develop an ontology that will characterize better the context and its various parameters.

## Acknowledgements

This work has been done within the context of the ITEA UseNet Project.